\documentstyle[12pt]{article}
\def\beq{\begin{equation}}
\def\eeq{\end{equation}}
\def\bea{\begin{eqnarray}}
\def\eea{\end{eqnarray}}
\def\bq{\begin{quote}}
\def\eq{\end{quote}}

\def\gappeq{\mathrel{\rlap {\raise.5ex\hbox{$>$}}
{\lower.5ex\hbox{$\sim$}}}}

\def\lappeq{\mathrel{\rlap{\raise.5ex\hbox{$<$}}
{\lower.5ex\hbox{$\sim$}}}}

\begin{document}
\topmargin -0.5cm
\oddsidemargin -0.3cm
\pagestyle{empty}
\begin{flushright}
{HIP-1999-09/TH}
\end{flushright}
\vspace*{5mm}
\begin{center}
{\bf New Magnetically Ordered  State in the Electron Plasma of Metals} \\
\vspace*{1.5cm} 
{\bf Christofer Cronstr\"{o}m}$^{*)}$ \\
\vspace{0.3cm}
Division of Theoretical Physics, Physics Department\\
P. O. Box 9, FIN-00014 University of Helsinki, Finland \\
\vspace{0.5cm}
and\\
\vspace{0.5cm}
{\bf Milan Noga}$^{\dagger) }$  \\
\vspace{0.3cm}
Helsinki Institute of Physics, P. O. Box 9, FIN-00014 University of Helsinki, Finland\\
\vspace*{1cm}  
{\bf ABSTRACT} \\ \end{center}
\vspace*{5mm}
\noindent
Within the simplest model of metals, namely a gas of electrons with Coulomb interactions,  
in the presence of a uniform background of positive charge to enforce electric neutrality 
of the system, we have derived a mechanism, by which the Coulomb interaction between the 
electrons generates a new kind of magnetism. The ground state of 
the metal is represented by a magnetically ordered state described by a non-local
magnetic field. This non-local magnetic field does not produce spin polarisation of
electrons, but induces a special long range correlation between electrons of opposite 
spin. This mechanism results in a theoretical value for the binding energy per 
electron, which is more than twice the corresponding value for the unmagnetised state of the metal. 
The new magnetic order proposed and analysed theoretically here, can in principle be 
experimentally tested.

\vspace*{3cm} 
\noindent 


\noindent
$^{*)}$ e-mail: Christofer.Cronstrom@Helsinki.fi\\
$^{\dagger)}$ e-mail: Milan.Noga@fmph.uniba.sk.\\
Permanent address: Department of Theoretical Physics, Comenius University, Mlynska Dolina\\ 
84215 Bratislava and Department of Physics, Faculty of Natural Sciences at the University of Constantinus 
Philosopher in Nitra, Slovak Republic.

\vfill\eject


\setcounter{page}{1}
\pagestyle{plain}

\section{Introduction}  

In considering many body systems, especially condensed matter, using Quantum Field Theory, it is
frequently advantageous to use  functional integral methods rather than canonical operator methods.
A particular benefit of functional integral methods is the possibility to eliminate  certain
fermionic functional integrals by means of a Grasmannian change of integration variables, 
and to obtain in stead  bosonic integrals involving new bosonic field variables. These bosonic
fields can then be associated with existing macroscopic structures or phases appearing in
interacting fermionic systems. To explain this in principle, consider a system of interacting 
fermions described by a given action $S_{f}[\psi^{*}, \psi]$, where $\psi^{*}(\tau)$ and
$\psi(\tau)$ are anticommuting (Grassmann-odd) variables, and $\tau$ is a parameter in the interval
$(0, \beta)$, with $\beta$ the inverse temperature. 

All the termodynamical properties of the system in its equilibrium state are obtained from the 
partition function $Z$, which is expressed as a functional integral as follows,
\begin{equation}
Z = \int D(\psi^{*}, \psi) \exp \left \{ - S_{f}[\psi^{*}, \psi] \right \},
\label{eq:partfZ}
\end{equation}
where $D(\psi^{*}, \psi)$ is a standard measure over Grassmann variables \cite{Vasiliev},
\cite{Kleinert}, \cite{Popov}. Making a change of integration variables in (\ref{eq:partfZ})
known as the Hubbard-Stratonovich transformation, or more generally a change of variables 
which introduces Grassmann-even, i.e. bosonic variables $B(\tau)$, one obtains a new functional
integral representation  \cite{Kleinert}, \cite{Popov} of the partition function (\ref{eq:partfZ})
\begin{equation}
Z = \int D(B, \psi^{*}, \psi) \exp \left \{ - S_{bf}[B, \psi^{*}, \psi] \right \}.
\label{eq:HSpartfZ}
\end{equation}
The second representation (\ref{eq:HSpartfZ}) is exact but not nesessarily useful or interesting,
unless one manages to choose the bosonic field variables in some particularly convenient way.
This depends naturally on the functional form of the original fermionic action 
$S_{f}[\psi^{*}, \psi]$. There may exist several different actions $S_{bf}[B, \psi^{*}, \psi]$,
corresponding to the same $S_{f}[\psi^{*}, \psi]$. Under  certain circumstances one may carry out 
the  integration over the fermionic variables in (\ref {eq:HSpartfZ}) explicitly and exactly,
obtaining an expression involving a bosonic field only,
\begin{equation} 
Z = \int D(B) \exp \left \{ - S_{eff}[B] \right \},
\label{eq:Zeff}
\end{equation}    
where $S_{eff}[B]$ is an effective action functional involving only bosonic variables $B$.
The functional integral (\ref {eq:Zeff}) can in general not be carried out explicitly, but may
be very convenient as a a starting point for approximate calculations. 

We will presently use the machinery presented in general terms so far, to analyse the simplest model 
of metals, namely an electron gas with Coulomb interaction between electrons, in the presence of 
a uniform background of positive charge, which is used in order to enforce the electric neutrality 
of the system. However it is convenient to discuss the approximate evaluation of the functional 
integral (\ref {eq:Zeff}) in general terms before considering detailed calculations. 

The approximate evaluation of (\ref {eq:Zeff}) goes in general as follows. One expands the action
$S_{eff}[B]$ as a power series in the fluctuations ${\cal B} = B - \overline{B}$ around its stationary 
value $S_{eff}[\overline{B}]$. The mean field $\overline{B}$ is a solution to the variational Euler-Lagrange 
equations,
\begin{equation}
\delta \,S_{eff}[B] = 0.
\label{eq:EuLa}
\end{equation}
Writing
\begin{equation}
S_{eff}[B] =  S_{eff}[\overline{B}] + S_{fl}[{\cal B}]
\label{eq:sadl1}
\end{equation}
one can then write,
\begin{equation}
Z = Z_{0} < \exp \left \{ - S_{fl}[{\cal B}] \right \}>_{0}\;, Z_{0} = \exp \left \{-S_{eff}[\overline{B}]\right \},
\label{eq:sadl2}
\end{equation}
where the symbol $<...>_{0}$ stands for average value with respect to $S_{eff}[\overline{B}]$.  In this
method, the saddle point contribution is treated exactly, while the effects of the fluctuations
around $\overline{B}$, are treated perturbatively, to the desired accuracy.

If the replacement of a fermionic integral with a bosonic one is succesfully carried out along 
the lines indicated above, then the bosonic fields $\overline{B}$ ought to describe  macroscopic 
structures, or phases, of the interacting fermionic system, such as ferromagnetism, 
superconductivity and so on, for interacting electron systems. The effects of fluctuations
\begin{equation}
< \exp \left \{ - S_{fl}[{\cal B}] \right \}>_{0}
\label{eq:fluct2}
\end{equation}   
ought to give rise to small perturbative corrections only, to the underlying structure of the 
system described by the field $\overline{B}$. If one can ignore effects of fluctuations, then the 
partition function $Z$ of the interacting fermion system is approximated by $Z_{0}$, and one says 
that the system is described in the mean field approximation represented by the field $\overline{B}$.

It was already mentioned above, that in general there may exist several realisations of fields
$B$ by means of which one transforms the functional integral (\ref {eq:partfZ}) into the
equivalent form (\ref {eq:HSpartfZ}). Hence, in the same fermionic system, one may in principle
obtain several macroscopic structures, or phases, described by a set of mean filds $\overline{B}^{(i)}$, where
the index $(i)$ enumerates the different macroscopic structures that can appear in the considered
fermionic system, under given external conditions. With each mean field $\overline{B}^{(i)}$ one then associates
a corresponding perturbation expansion, given by the relation (\ref {eq:sadl2}), for $\overline{B} = \overline{B}^{(i)}$.
If there are several mean fields $\overline{B}^{(i)}, i = 1,2,...$ one may ask which of these is 
actually realised in the system at hand. There is a nearly obvious answer to this; the structure
$\overline{B}^{(i)}$ associated with the lowest energy under given external conditions is realised. It may of
course happen that the external conditions do not lead to a unique structure $\overline{B}^{(i)}$, in which
case we have a coexistence of macroscopic phases.

\section{The simplest model of metals}

Let us now return to the physical problem of interest, namely the problem of the ground state in 
the simplest model of metals \cite{Gell-Mann}, which we analyse using the general method outlined 
above. The problem in question is ancient \cite{Bloch}, \cite{Fetter-W}, and concerns more precisely 
the question of whether the genuine ground state of the metal is an unmagnetised state or a 
ferromagnetic state.

The model in question consists of a large number $N$ of electrons (an electron gas) in a finite but 
large volume V, interacting through the ordinary Coulomb interaction, in the presence of a uniform 
background of positive charge determined so that the system is electrically neutral. The system is 
described by the following Hamiltonian,
\begin{equation}
H = H_{0} + H_{int} - C,
\label{eq:Hammod}
\end{equation}
where
\begin{equation}
H_{0} = \int_{V} d^{3}{\bf x}\sum_{\alpha} \left [ - \frac{\hbar^{2}}{2m}\psi^{\dagger}_{\alpha}({\bf x})\nabla^{2}\psi_{\alpha}({\bf x}) \right ]
\label{eq:Hzero}
\end{equation}
describes the free electron gas, and
\begin{equation}
H_{int} = \frac{e^{2}}{2}\sum_{\alpha, \alpha'}\int_{V}\int_{V} \frac{ d^{3}{\bf x} d^{3}{\bf x}'}
{\mid {\bf x} - {\bf x}' \mid } 
\left \{\psi^{\dagger}_{\alpha'}({\bf x}')\psi^{\dagger}_{\alpha}({\bf x})\psi_{\alpha}({\bf x})\psi_{\alpha'}({\bf x}') \right \}
\label{eq:Hint}
\end{equation}
describes the Coulomb interaction between the electrons. Here the quantities $\psi^{\dagger}_{\alpha}$
and $\psi_{\alpha}$ are the creation and annihilation operators, respectively, of electrons with spin
component $\alpha$. Finally, the constant $C$ in the 
expression (\ref {eq:Hammod}) is the following,
\begin{equation}
C = \frac{e^{2}}{2}\eta^{2}\int_{V}\int_{V} \frac{ d^{3}{\bf x} d^{3}{\bf x}'}{\mid {\bf x} - {\bf x}' \mid },
\label{eq:Charge}
\end{equation}
where $e\eta$ is the density of uniformly distributed positive charge related to the average density of the electron charge as follows,
\begin{equation}
e\eta =  e\sum_{\alpha} < \psi^{\dagger}_{\alpha}({\bf x})\psi_{\alpha}({\bf x}) >.
\label{eq:defeta}
\end{equation}
Incorporating the constant $C$ as defined in (\ref {eq:Charge}) above in the total Hamiltonian
(\ref {eq:Hammod}), makes the system electrically neutral, as it should be.

The ground state energy $E_{0}$ of the unmagnetised state in the model above has been derived in
perturbation theory \cite{Gell-Mann}, \cite{Wigner1}, \cite{Macke}, \cite{Pines}, with  $H_{0}$ as 
the unperturbed Hamiltonian , and expressed as a sum of three terms,
\begin{equation}
E_{0} = E_{kin.} + E_{exch.} + E_{corr.}
\label{eq:GrE1}
\end{equation}
The expression for the leading term $E_{kin.}$ in (\ref {eq:GrE1}) is as follows,
\begin{equation}
E_{kin.} = \frac{e^{2}N}{2a_{0}}\,\frac{3}{5} \left (\frac{9\pi}{4} \right )^{\frac{2}{3}} \frac{1}{r_{s}^{2}} \approx \frac{e^{2}N}{2a_{0}}\,\frac{2.21}{r_{s}^{2}}, 
\label{eq:Ekin}
\end{equation}
where $N$ is the total number of electrons in the system, $a_{0}$ is the Bohr radius, 
$r_{s} = r_{0}/a_{0}$ and $r_{0}$ is a parameter related to the volume per electron, 
i.e. $\frac{4}{3}\pi r_{0}^{3} = V/N$.

The term $E_{exch.}$ in (\ref {eq:GrE1}) is the exchange energy, i.e. the expectation value of
$H_{int.}$ in the ground state of a free electron gas, and has the following form,
\begin{equation}
E_{exch.} = - \frac{e^{2} N }{2a_{0}}\,\frac{3}{2\pi} \left (\frac{9\pi}{4} \right )^{\frac{1}{3}} \frac{1}{r_{s}} \approx - \frac{e^{2} N }{2a_{0}}\,\frac{0.916}{r_{s}}.
\label{eq:Eexch}
\end{equation}
The last term in the expression (\ref {eq:GrE1}), is the so-called correlation energy, a term introduced by 
Wigner \cite{Wigner1}. This term was calculated by Gell-Mann and Brueckner to the accuracy $e^{4}$,
by summing an infinite number of so-called ring diagrams, with the result,
\begin{equation}
E_{corr.} = \frac{e^{2} N }{2a_{0}}\,\left [\frac{2}{\pi^{2}}(1-\log2)\log r_{s} - 0.096 \right ] \approx \frac{e^{2} N }{2a_{0}}\,\left [0.0622 \log r_{s} - 0.096 \right ].
\label{eq:Ecorr}
\end{equation}

It was pointed out by F. Bloch already in 1929 \cite{Bloch} that the ferromagnetic state of the 
electron plasma in metals could have a lower energy than that of the unmagnetised state. By considering
a polarised gas of electrons with $N_{+}$ electrons in the spin-up state, and $N_{-}$ electrons in the 
spin-down state and defining the polarisation $\xi = (N_{+} - N_{-})/N$, where $N = N_{+} + N_{-}$, 
one finds the following expression for the ground state energy to the first order in the interaction 
potential,
\begin{eqnarray}
\label{eq:Emag}
E_{mag} & = & E_{kin.} + E_{exch.}\\
&  = & \frac{e^{2}N}{2a_{0}}\, \frac{3}{10} \left (\frac{9\pi}{4}\right )^{\frac{2}{3}} \left [(1+\xi)^{\frac{5}{3}} + (1-\xi)^{\frac{5}{3}}\right ]\frac{1}{r_{s}^{2}}\nonumber \\
&  & - \frac{e^{2}N}{2a_{0}}\, \frac{3}{4\pi} \left (\frac{9\pi}{4}\right )^{\frac{1}{3}} \left [(1+\xi)^{\frac{4}{3}} + (1-\xi)^{\frac{4}{3}}\right ]\frac{1}{r_{s}}. \nonumber
\end{eqnarray}
One readily infers from the expression (\ref {eq:Emag}) that the ferromagnetic state, with
$\xi = 1$, has a lower energy than the unmagnetised state with $\xi = 0$, provided that
\begin{equation}
r_{s} > \frac{2\pi}{5}\,\left (\frac{9\pi}{4} \right )^{\frac{1}{3}} [2^{\frac{1}{3}} + 1] \approx 5.45.
\label{eq:ineq}
\end{equation}
This problem is also discussed in the textbook by Fetter and Walecka \cite{Fetter-W}.

In the relation (\ref {eq:Emag}) the polarisation $\xi$ is a free parameter subject to variation
in the interval $[0, 1]$. The expression (\ref {eq:Emag}) represents, in fact, only the energy of the 
electrons in a uniform magnetic field, which gives rise to the polarisation of these electrons. 
If the  electrons become spontaneously polarised, then there should appear a spontaneously 
self-organised magnetic field ${\bf H}$ in the system, which makes a positive contribution $E_{field}$ 
to the total energy of the system. Then the total energy of the ferromagnetic state $E_{ferro}$ 
would be the following,
\begin{equation}
E_{ferro} = E_{mag} + E_{field}
\label{eq:ferroE}
\end{equation}
There is the possibility, that the energy $E_{ferro}$ may be lower than the energy of the
unmagnetised state given to lowest order by the sum of (\ref {eq:Ekin}) and (\ref {eq:Eexch}).
Thus the ferromagnetic state can in principle be a genuine ground state of the system for certain 
external conditions.

In \cite{Wigner2} and \cite{Pines} doubts were expressed concerning the possibility of having a
ferromagnetic state as a genuine ground state of the Hamiltonian (\ref {eq:Hammod}) in view of 
the fact, among other things, that the contribution $E_{field}$ was neglected in the considerations 
based on (\ref {eq:Emag}).

Independently of arguments for or against the possibility of a ferromagnetic ground state for the 
system described by the Hamiltonian (\ref {eq:Hammod}), the Bloch considerations based on 
(\ref {eq:ferroE}) pose a challenge in electron plasma theory. The challenging task is to derive
a mechanism by means of which the Coulomb interaction between electrons  in the model described by
(\ref {eq:Hammod}), can generate magnetism. To the best of our knowledge, this has been  an unsolved 
problem so far. Below we give a solution to this problem, i.e. we show, using the functional 
methods discussed in general terms in the Introduction, that there is a kind of magnetic field that 
arises due to the Coulomb interaction, and that the corresponding magnetic state under certain
conditions is a genuine ground state of the system under consideration. 

\section{Magnetic ordering in the electron plasma}

The grand canonical partition function $Z$ corresponding to the Hamiltonian
(\ref {eq:Hammod}) is given by the functional integral (\ref {eq:partfZ}) with the 
following action
\begin{eqnarray}
\label{eq:actSf}
\lefteqn{S_{f}[\psi^{*}, \psi]  = } \\
& & \int_{0}^{\beta} d\tau \sum_{\alpha} \int d^{3}{\bf x}
\left [ \psi_{\alpha}^{*}({\bf x}, \tau) \frac{\partial}{\partial \tau}\psi_{\alpha}({\bf x}, \tau) - \frac{\hbar^{2}}{2m}\psi_{\alpha}^{*}({\bf x}, \tau )\nabla^{2} \psi_{\alpha}({\bf x}, \tau ) - \mu \psi_{\alpha}^{*}({\bf x}, \tau ) \psi_{\alpha}({\bf x}, \tau ) \right ] \nonumber \\ 
&  & + \frac{e^{2}}{2}\int_{0}^{\beta} d\tau \sum_{\alpha, \alpha'}\int \int \frac{ d^{3}{\bf x} d^{3}{\bf x}'}{\mid {\bf x} - {\bf x}' \mid } 
\left [\psi^{*}_{\alpha'}({\bf x}', \tau )\psi^{*}_{\alpha}({\bf x}, \tau )\psi_{\alpha}({\bf x}, \tau ) \psi_{\alpha'}({\bf x}', \tau) \right ] - \beta C, \nonumber
\end{eqnarray}
where the electron  field variables $\psi_{\alpha}^{*}({\bf x}, \tau )$ and
$\psi_{\alpha}({\bf x}, \tau)$ are Grassmann  variables enumerated by the spin index $\alpha$,
the space coordinate ${\bf x}$ and the "time" $\tau$, and $\mu$ is the chemical potential. 

The electron  field variables satisfy the antiperiodic boundary conditions,
\begin{equation}
\psi_{\alpha}^{*}({\bf x}, \tau + \beta ) = - \psi_{\alpha}^{*}({\bf x}, \tau )\,,\; \psi_{\alpha}({\bf x}, \tau + \beta ) = - \psi_{\alpha}({\bf x}, \tau ).
\label{eq:antipbc}
\end{equation}

\subsection{The unmagnetised electron plasma}

We first review one known \cite{Kleinert}, \cite{Popov} way of introducing bosonic variables
in the fermionic system described in terms of the functional integral (\ref {eq:partfZ}). This consists
of introducing an electric scalar potential $\phi({\bf x}, \tau)$ , in the present case when the 
action is given by the expression (\ref {eq:actSf}) above.

As mentioned in the Introduction,  there may in general be more than one way of transforming the
functional integral (\ref {eq:partfZ}) into a functional integral (\ref {eq:HSpartfZ}) involving
also bosonic (Grassmann even) variables. We consider an alternative to the procedure explained below 
in the next subsection.

The partition function $Z$ involving the electric scalar potential is the following,
\begin{equation}
Z =  \frac{1}{Z_{0 \phi}}\int D(\phi, \psi^{*}, \psi) \exp \left \{ - S_{bf}[\phi, \psi^{*}, \psi] \right \},
\label{eq:scalpot}
\end{equation}
where
\begin{eqnarray}
\label{eq:scalact}
S_{bf}[\phi, \psi^{*}, \psi] & = &  \frac{1}{8\pi }\int_{0}^{\beta} d\tau \int d^{3} {\bf x}\; \left (\nabla \phi({\bf x}, \tau)\right )^{2} \\
& & + \int_{0}^{\beta} d\tau \int d^{3} {\bf x}\;\sum_{\alpha} \psi_{\alpha}^{*}({\bf x}, \tau) \frac{\partial}{\partial \tau}\psi_{\alpha}({\bf x} , \tau) \nonumber\\
& & + \int_{0}^{\beta} d\tau \int d^{3} {\bf x} \sum_{\alpha} \left [ - \frac{\hbar^{2}}{2m}\psi_{\alpha}^{*}({\bf x}, \tau )\nabla^{2} \psi_{\alpha}({\bf x}, \tau ) - \mu \psi_{\alpha}^{*}({\bf x}, \tau ) \psi_{\alpha}({\bf x}, \tau )\right ] \nonumber \\
& & + ie \int_{0}^{\beta} d\tau \int d^{3} {\bf x}\; \phi({\bf x}, \tau) \left [ \eta - \sum_{\alpha} \psi_{\alpha}^{*}({\bf x}, \tau ) \psi_{\alpha}({\bf x}, \tau )\right ]. \nonumber
\end{eqnarray}
In Eq. (\ref {eq:scalact}) above, the quantity $e\eta$ is a uniform positive charge density and $Z_{0\phi}$ is
defined by the following functional integral,
\begin{equation}
Z_{0\phi} = \int D(\phi) \exp \left \{- \frac{1}{8\pi }\int_{0}^{\beta} d\tau \int d^{3} {\bf x}\, [\nabla \phi({\bf x}, \tau)]^{2} \right \}.
\label{eq:Zophi}
\end{equation}
The quantiy $\phi({\bf x}, \tau)$ introduced above, is a bosonic, i.e. Grassmann even, integration
variable, which satisfies the following periodic boundary conditions,
\begin{equation}
\phi({\bf x}, \tau + \beta) = \phi({\bf x}, \tau)
\label{eq:bcphi}
\end{equation}
In the action (\ref {eq:scalact}) one can recognize a contribution from the energy density of an electric field
${\bf E}({\bf x}, \tau) = - \nabla \phi({\bf x}, \tau)$, and from a system of electrons interacting
through a scalar potential $\phi({\bf x}, \tau)$. Performing the integration over the variable
$\phi({\bf x}, \tau)$ in the functional integral (\ref {eq:scalpot}) one recovers indeed the
functional integral (\ref {eq:partfZ}) with the action $S$ given by Eq. (\ref {eq:actSf}). However, one may also 
first integrate over the variables $\psi^{*}$ and $\psi$ in Eq. (\ref {eq:scalpot}). One then 
obtains,
\begin{equation}
Z = \frac{1}{Z_{0\phi}} \int D[\phi] \exp 
\left \{ - \int_{0}^{\beta} d\tau \int d^{3} {\bf x}\;\left [\frac{1}{8\pi}[\nabla \phi({\bf x}, \tau)]^{2} + ie \eta \phi({\bf x}, \tau) \right ]\, det\;{\cal F}(\phi) \right \},
\label{eq:Zbos}
\end{equation}
wherere ${\cal F}(\phi)$ is a functional matrix, the matrix elements of which are as follows,
\begin{equation}
<{\bf x}', \tau', \alpha' \mid {\cal F}(\phi) \mid {\bf x}, \tau , \alpha >\; = 
\left [\frac{\partial}{\partial \tau} - \frac{\hbar^{2}}{2m} \nabla_{\bf x}^{2} - \mu -ie \phi({\bf x}, \tau ) \right ]\, \delta(\tau' - \tau) \,\delta({\bf x}' - {\bf x})\, \delta_{\alpha' \alpha}.
\label{eq:Fm-elem}
\end{equation}
The expression (\ref {eq:Zbos}) defines a certain effective action $S_{eff}$. Evaluating the
functional integral (\ref {eq:Zbos}) approximatively, to the order $e^{4}$, using the general 
procedure given by Eqns. (\ref {eq:EuLa}) and (\ref {eq:sadl2}) above, one reproduces the 
expressions (\ref {eq:Ekin}), (\ref {eq:Eexch}) and (\ref {eq:Ecorr}) in the expression 
(\ref {eq:GrE1}) for the ground state energy $E_{0}$ of the unmagnetised electron plasma.

\subsection{The magnetised electron plasma}

We now come to the central issue in this paper, namely that there is another way of recognising
bosonic variables in the system described by the action (\ref {eq:actSf}), than the one described above, which 
utilised a scalar field $\phi({\bf x}, \tau )$. The novel transformation involves {\em bilocal}
fields.

Consider auxiliary commuting (Grassmann even) real variables $A_{\alpha'\alpha}({\bf x}', {\bf x}, \tau )$,
enumerated by spin indices $\alpha ,\alpha' = \pm$, by two space coordinates
${\bf x}'$ and ${\bf x}$ (${\bf x}' \neq {\bf x}$) and by a "time-variable" $\tau $. The
commuting variables $A_{\alpha'\alpha}({\bf x}', {\bf x}, \tau )$ are taken to be antisymmetric
with respect to a simultaneous transposition of space- and spin indices,
\begin{equation}
A_{\alpha'\alpha}({\bf x}', {\bf x}, \tau ) = - A_{\alpha\alpha'}({\bf x}, {\bf x}', \tau )
\label{eq:antisA}
\end{equation}
Furthermore, these variables are supposed to satisfy periodic boundary conditions,
\begin{equation}
A_{\alpha'\alpha}({\bf x}', {\bf x}, \tau + \beta ) = A_{\alpha'\alpha}({\bf x}', {\bf x}, \tau ) 
\label{eq:perbcA}
\end{equation}
The variables  $A_{\alpha'\alpha}$ can be regarded as matrix elements of a certain $2\times 2$ 
matrix, expressed with the aid of the $2\times 2$ unit matrix $\sigma^{0} = 1$ and the usual 
Pauli matrices $\sigma^{k}, ( k = 1,2,3)$, and a four-component quantity $B = (B_{0},B_{1}, B_{2}, B_{3})$ 
as follows,
\begin{equation}
A_{\alpha'\alpha}({\bf x}', {\bf x}, \tau ) = \sum_{\ell=0}^{3} \sigma^{\ell}_{\alpha'\alpha} B_{\ell}({\bf x}', {\bf x}, \tau )
\label{eq:Arepr}
\end{equation}

The symmetry conditions (\ref {eq:antisA}) satisfied by the real quantities $A_{\alpha'\alpha}$
imply the following reality and symmetry conditions for the four-component quantity $B$,
\begin{equation}
Im \;B_{k}({\bf x}', {\bf x}, \tau ) \equiv 0, \;B_{k}({\bf x}', {\bf x}, \tau ) = -  B_{k}({\bf x}, {\bf x}', \tau )\;\;, k = 0,1,3.
\label{eq:syprB013}
\end{equation}
The second component $B_{2}$ is purely imaginary, and symmetric under the interchange of ${\bf x}$
and ${\bf x}'$,
\begin{equation}
Re \;B_{2}({\bf x}', {\bf x}, \tau ) \equiv 0,  \; B_{2}({\bf x}', {\bf x}, \tau )  = + \, B_{2}({\bf x}, {\bf x}', \tau )
\label{eq:syprB2}
\end{equation}

The real variables $A_{\alpha'\alpha}({\bf x}', {\bf x}, \tau )$ are used to define an action
$S_{0b}$ as follows,
\begin{eqnarray}
\label{eq:Brepr}
S_{0b}[B] & = &\frac{1}{4e^{2}} \int_{0}^{\beta}d\tau \int\int d^{3}{\bf x}d^{3}{\bf x}' \mid {\bf x} - {\bf x}'\mid \sum_{\alpha'\alpha} A^{2}_{\alpha'\alpha}({\bf x}', {\bf x}, \tau ) \\
& = & \frac{1}{2e^{2}} \int_{0}^{\beta} d\tau \int\int d^{3}{\bf x}d^{3}{\bf x}'\mid {\bf x} - {\bf x}'\mid \sum_{k=0}^{3} B_{k}^{\dagger}({\bf x}', {\bf x}, \tau ) B_{k}({\bf x}', {\bf x}, \tau ), \nonumber
\end{eqnarray}
and the corresponding partition function,
\begin{equation}
Z_{0b} = \int D(B) \exp \left \{ - S_{0b}[B] \right \}.
\label{eq:Z0b[B]}
\end{equation}
We wish to evaluate the  partition function (\ref {eq:partfZ}) with the action $S_{f}$ given by
Eq. (\ref {eq:actSf}). To this end we multiply Eq. (\ref {eq:partfZ}) with the evident identity, 
\begin{equation}
1 = \frac{1}{Z_{0b}} \int D(B) \exp \left \{ - S_{0b}[B] \right \}.
\label{eq:iden}
\end{equation}
This results in the following expression for the partition function $Z$,
\begin{equation}
Z =  \frac{1}{Z_{0b}}\int D(B) D(\psi^{*}, \psi) \exp \left \{ - S_{0b}[B] - S_{f}[\psi^{*}, \psi] \right \}
\label{eq:finZ}
\end{equation}
In the expression (\ref {eq:finZ}) we finally make the following change of integration variables,
\begin{eqnarray}
\label{eq:shift}
A_{\alpha'\alpha}({\bf x}', {\bf x}, \tau ) & \rightarrow & A'_{\alpha'\alpha}({\bf x}', {\bf x}, \tau ) =  A_{\alpha'\alpha}({\bf x}', {\bf x}, \tau ) + \\
& & + \frac{ie^{2}}{\mid {\bf x}' - {\bf x} \mid } [\psi^{*}_{\alpha'}({\bf x}', \tau)\psi_{\alpha}({\bf x}, \tau) - \psi^{*}_{\alpha}({\bf x}, \tau)\psi_{\alpha'}({\bf x}', \tau)]. \nonumber
\end{eqnarray}
 
The reason for the shift of integration variable described above, is simply that by evaluating the
functional integral (\ref {eq:finZ}) in the manner indicated, one cancels the Coulomb interaction 
term in the expression (\ref {eq:actSf}) for $S_{f}[\psi^{*}, \psi]$ and is left with the following simple 
expression,
\begin{equation}
Z = \frac{1}{Z_{0b}} \int D(B, \psi^{*}, \psi) \exp \left \{- S_{bf}[B, \psi^{*}, \psi] \right \}
\label{eq:ZpsiB}
\end{equation}
with an action $S_{bf}[B, \psi^{*}, \psi]$ given by the following expression,
\begin{eqnarray}
\label{eq:Sbfin}
S_{bf}[B, \psi^{*}, \psi] & = & S_{0b}[B] + \int_{0}^{\beta} \int d^{3}{\bf x} \psi^{*}({\bf x}, \tau)\left [ \frac{\partial}{\partial \tau} - \frac{\hbar^{2}}{2m} \nabla^{2} - \mu \right ]\psi({\bf x}, \tau) \\
&  & + i\int_{0}^{\beta} \int \int d^{3}{\bf x} d^{3}{\bf x}'\psi^{*}({\bf x}', \tau)\;\left (\sum_{k=0}^{3}\sigma^{k} B_{k}({\bf x}', {\bf x}, \tau) \right ) \;\psi({\bf x}, \tau) - \beta C \nonumber
\end{eqnarray}
In the expression (\ref {eq:Sbfin}) we have used ordinary two-component notation,
\begin{equation}
\psi = \left (\begin{array}{c}
                          \psi_{+} \\ \psi_{-}
                          \end{array} \right ) \;,\; \psi^{*} = (\psi^{*}_{+}, \psi^{*}_{-})
\label{eq:spinot}
\end{equation}
for the Grassmann variables $\psi^{*}({\bf x}, \tau)$ and $\psi({\bf x}, \tau)$, respectively.

The action (\ref {eq:Sbfin}) apparently describes a system of electrons interacting through a bilocal scalar 
potential field $B_{0}({\bf x}',{\bf x}, \tau)$ and a bilocal magnetic field 
${\bf B}({\bf x}',{\bf x}, \tau)$. The special symmetry properties of these field variables under 
interchange of space coordinates are given in Eqns. (\ref {eq:syprB013}) and (\ref {eq:syprB2}).

The formal transformations (\ref {eq:finZ}) and (\ref {eq:shift}) leading from the original action
(\ref {eq:actSf}) to the action (\ref {eq:Sbfin}), which describes a system of electrons interacting through 
a bilocal field $B({\bf x}, {\bf x}', \tau)$, define a mechanism through which magnetic 
interactions can be generated from the Coulomb interaction. This mechanism is an alternative 
to the transformation involving only an electric scalar potential $\phi({\bf x}, \tau)$ 
which resulted in the 
action $S_{bf}$ given in Eq. (\ref {eq:scalact}), by means of which one could reproduce the  
expressions (\ref {eq:Ekin}), (\ref {eq:Eexch}) and (\ref {eq:Ecorr}) in the formula
(\ref {eq:GrE1}) for the ground state energy $E_{0}$ of the unmagnetised electron plasma.
Clearly, the self-consistency of the mechanism described above will have to be ascertained. We
return to this question presently, and continue now with a straightforward analysis
based on the action (\ref {eq:Sbfin}), which is bilinear in the fields $\psi^{*}$ and $\psi$. 
This circumstance makes it possible to carry out the functional integrations
over $\psi^{*}$ and $\psi$ in the expression (\ref {eq:ZpsiB}) exactly and explicitly, with the
result,
\begin{equation}
Z = \frac{e^{\beta C}}{Z_{0b}} \int D[B] \exp \left \{ - S_{0b}[B] \right \}\;det \;{\cal F}(B),
\label{eq:Zfin2}
\end{equation}
where ${\cal F}(B)$ is a functional matrix, the matrix elements of which can be written as follows,
\begin{eqnarray}
\label{eq:functmatr}
<{\bf x}', \tau', \alpha' \mid {\cal F}(B) \mid {\bf x}, \tau, \alpha> & = &  \left [\frac{\partial}{\partial \tau} - \frac{\hbar^{2}}{2m} \nabla^{2} - \mu \right ]\; \delta_{\alpha'\alpha}\; \delta(\tau' - \tau)\; \delta({\bf x}'- {\bf x}) \\
& & + i\;\sum_{\ell=0}^{3}\sigma^{\ell}_{\alpha'\alpha} B_{\ell}({\bf x}', {\bf x}, \tau) \;\, \delta(\tau' - \tau) \nonumber
\end{eqnarray}
The Dirac $\delta$-function $\delta(\tau' - \tau)$ which enters in the relation (\ref {eq:functmatr}) above,
is here defined in terms of its spectral representation,
\begin{equation}
\delta(\tau' - \tau) = \frac{1}{\beta} \sum_{\nu = -\infty}^{+\infty} \exp i\omega_{\nu}(\tau' - \tau),\; \omega_{\nu} = \frac{\pi}{\beta}(2\nu + 1),
\label{eq:deltatau}
\end{equation}
where the summation index $\nu$ runs through all integer values, and the quantities $\omega_{\nu}$
are the so-called Matsubara frequencies.

Employing the relation
\begin{equation}
det\;{\cal F}(B) = \exp \left \{Tr \log {\cal F}(B) \right \}
\label{eq:detxtr}
\end{equation}
we obtain the final result for the partition function $Z$ in Eq. (\ref {eq:Zfin2}),
\begin{equation}
Z = \frac{1}{Z_{0b}} \int D(B) \exp \left \{ - S_{eff}[B] \right \},
\label{eq:Zfin3}
\end{equation}
where the effective action $S_{eff}[B]$ is as follows,
\begin{equation}
S_{eff}[B] = S_{0b}[B] - \sum_{\alpha} \int_{0}^{\beta} d\tau \int d^{3}{\bf x}<{\bf x}, \tau, \alpha \mid \log {\cal F}(B) \mid {\bf x}, \tau, \alpha> - \beta C.
\label{eq:FinSeff}
\end{equation}
The equation (\ref {eq:FinSeff}) will be the starting point for an approximate evaluation of the 
partition function $Z$ under consideration. This problem will be considered in the next subsection.

\subsection{The saddle point equations}

We now look for the extremals of the effective action (\ref {eq:FinSeff}), i.e. for solutions
to the equations,
\begin{equation}
\label{eq:varSeff}
\delta \,S_{eff}[B] = 0.
\end{equation}
It is reasonably simple to evaluate the variations involved in Eq. (\ref {eq:varSeff}). Taking in
addition into account the  reality and symmetry conditions (\ref {eq:syprB013}) and (\ref {eq:syprB2}) 
above, one obtains the following four equations from the conditions (\ref {eq:varSeff}). For 
$\ell = 0,1,3$ the equations in question are, 
\begin{eqnarray}
\label{eq:vareqB013}
\lefteqn{\frac{1}{e^{2}} \mid {\bf x} - {\bf x}' \mid B_{\ell}({\bf x}, {\bf x}', \tau)  = } \\
& & = \frac{i}{2}\sum_{\alpha \alpha'} \sigma^{\ell}_{\alpha \alpha'} \left \{ <{\bf x}', \tau, \alpha' \mid {\cal F}^{-1}(B) \mid {\bf x}, \tau, \alpha > - <{\bf x}, \tau, \alpha' \mid {\cal F}^{-1}(B) \mid {\bf x}', \tau, \alpha > \right \}  \nonumber 
\end{eqnarray}
For $\ell = 2$ one gets,
\begin{eqnarray}
\label{eq:vareqB2}
\lefteqn{\frac{1}{e^{2}} \mid {\bf x} - {\bf x}' \mid B_{2}({\bf x}, {\bf x}', \tau)   = } \\
& & = - \frac{i}{2} \sum_{\alpha \alpha'} \sigma^{2}_{\alpha \alpha'} \left \{ <{\bf x}', \tau, \alpha' \mid {\cal F}^{-1}(B) \mid {\bf x}, \tau, \alpha > + <{\bf x}, \tau, \alpha' \mid {\cal F}^{-1}(B) \mid {\bf x}', \tau, \alpha > \right \}  \nonumber
\end{eqnarray}

Using the relation ({\ref {eq:functmatr}) one finds readily that the quantity $<{\bf x}', \tau', \alpha' \mid {\cal F}^{-1}(B) \mid {\bf x}'', \tau '', \alpha''>$
satisfies the following equation,
\begin{equation}
\label{eq:invF(B)}
\sum_{\alpha'} \int d^{3}{\bf x}' {\cal O}({\bf x}, \alpha; {\bf x}', \alpha') <{\bf x}', \tau', \alpha' \mid {\cal F}^{-1}(B) \mid {\bf x}'', \tau '', \alpha''> = \delta(\tau' - \tau'') \delta^{3}({\bf x} - {\bf x}'') \delta_{\alpha \alpha''}
\end{equation}
where the operator ${\cal O}$ is the following,
\begin{equation}
\label{eq:operO}
{\cal O}({\bf x}, \alpha; {\bf x}', \alpha') = \left \{ \left [- \frac{\partial}{\partial \tau'} - \frac{\hbar^{2}}{2m} \nabla_{{\bf x}}^{2} - \mu \right ] \delta_{\alpha \alpha'} \delta^{3}({\bf x} - {\bf x}') + i \sum_{\ell = 0}^{3} \sigma_{\alpha \alpha'}^{\ell} B_{\ell}({\bf x}, {\bf x}',\tau') \right \}.
\end{equation}

The equations (\ref {eq:vareqB013}) and (\ref {eq:vareqB2}) together with the equation
(\ref {eq:invF(B)}) constitute a complicated set of nonlinear integral equations for the
saddle point configuration $B_{\mu}({\bf x}, {\bf x}', \tau),\; \mu = 0,1,2,3$. It remains to show 
that a nontrivial saddle point configuration exists. We shall not attempt here to discuss the most 
general solution possible of the set of equations (\ref {eq:vareqB013}), (\ref {eq:vareqB2}) and 
(\ref {eq:invF(B)}). Rather we confine our attention to the following special $\tau$-independent
{\em ansatz}, which we denote by  $\overline{B}$,
\begin{equation}
\overline{B}_{\ell}({\bf x}, {\bf x}', \tau) = 0, \; \ell = 0,1,3;\; \overline{B}_{2}({\bf x}, {\bf x}', \tau) = -iH(\mid {\bf x} - {\bf x}' \mid ),
\label{eq:Hansatz}
\end{equation}
where $H(\mid{\bf x}\mid)$ is a real-valued function to be determined. Inserting the ansatz
(\ref {eq:Hansatz}) in the equation (\ref {eq:invF(B)}), one finds that this equation, apart from a
factor $-1$, is nothing but the equation for the Green's function for an electron in a magnetic field
of variable magnitude $H$ in the direction of the $2$-axis. Under these circumstances it is appropriate to 
introduce the following notation,
\begin{equation}
G_{\alpha' \alpha''}({\bf x}' - {\bf x}'', \tau' -\tau''; H) \equiv - <{\bf x}', \tau', \alpha' \mid {\cal F}^{-1}(\overline{B}) \mid {\bf x}'', \tau '', \alpha''>.
\label{eq:defGreen}
\end{equation}
 
We  consider first the equation (\ref {eq:vareqB2}),
which becomes an equation for the determination of the function $H(\mid {\bf x} \mid)$ in the ansatz
(\ref {eq:Hansatz}). In order to analyse this equation, it is convenient to consider the Fourier
transform $\tilde{G}_{\alpha \alpha'}$ of the Green's function $G_{\alpha \alpha'}$ defined by the 
equation (\ref {eq:defGreen}),
\begin{equation}
\tilde{G}_{\alpha \alpha'}({\bf k}, \nu; H) \equiv \int_{0}^{\beta} d\tau \int_{V} d^{3}{\bf x}\; G_{\alpha  \alpha'}({\bf x}, \tau; H) \exp (-i\omega_{\nu}\tau + i{\bf k}\cdot {\bf x}).
\label{eq:FouriG}
\end{equation}
Using the equation (\ref {eq:invF(B)}) in the present case, one readily finds that
\begin{equation}
\sum_{\alpha'} \left \{ [i\omega_{\nu} - (\frac{\hbar^{2}{\bf k}^{2}}{2m} - \mu) ]\delta_{\alpha \alpha'} - \sigma^{2}_{\alpha \alpha'} \tilde{H}({\bf k}) \right \} \tilde{G}_{\alpha' \alpha''}({\bf k}, \nu; H) = \delta_{\alpha \alpha''},
\label{eq:EtildG}
\end{equation}
where 
\begin{equation}
\tilde{H}({\bf k}) = \int_{V} d^{3}{\bf x} H(\mid {\bf x} \mid) \exp (i{\bf k}\cdot {\bf x}).
\label{eq:FouriH}
\end{equation}
The Eq. (\ref {eq:EtildG}) has the following solution,
\begin{equation}
\tilde{G}_{\alpha  \alpha'}({\bf k}, \nu; H) = \frac{(i\omega_{\nu} - {\cal E}_{0}({\bf k}) + \mu)\,\delta_{\alpha \alpha'} + \sigma^{2}_{\alpha \alpha'}\tilde{H}({\bf k})}{(i\omega_{\nu} - {\cal E}_{+}({\bf k}) + \mu)(i\omega_{\nu} - {\cal E}_{-}({\bf k}) + \mu)},
\label{eq:soltildeG}
\end{equation}
where
\begin{equation}
{\cal E}_{0}({\bf k}) = \frac{\hbar^{2} {\bf k}^{2}}{2m}, \;{\cal E}_{\pm}({\bf k}) = \frac{\hbar^{2} {\bf k}^{2}}{2m} \pm \tilde{H}({\bf k}).
\label{eq:defcalE}
\end{equation}

In passing we note the symmetry of the Green's function in Eq. (\ref {eq:defGreen}) under the 
interchange of ${\bf x}'$ and ${\bf x}''$, which readily follows from the equation (\ref {eq:soltildeG}) 
above. From this symmetry follows that the ansatz (\ref {eq:Hansatz}) is {\em consistent}
with the remaining equations (\ref {eq:vareqB013}). We then continue with our analysis of Eq.
(\ref {eq:vareqB2}).

The equation (\ref {eq:vareqB2}) now finally boils down to the following,
\begin{equation}
\frac{1}{e^{2}}\mid {\bf x} \mid H(\mid {\bf x} \mid)  = - \sum_{\alpha \alpha'} \sigma^{2}_{\alpha \alpha'} G_{\alpha' \alpha}({\bf x}, 0; H),
\label{eq:fineqH}
\end{equation}
where
\begin{equation}
G_{\alpha' \alpha}({\bf x}, 0; H) =  \frac{1}{V\beta}  \sum_{{\bf k}}\,\lim_{N \rightarrow \infty}\sum_{\nu = -N}^{N} \tilde{G}_{\alpha' \alpha}({\bf k}, \nu; H) \exp (-i{\bf k}\cdot{\bf x}).
\label{eq:finGeq}
\end{equation}
Inserting the expression (\ref {eq:soltildeG}) in Eq. (\ref {eq:finGeq}) and using the well-known
series,
\begin{equation}
\lim_{N \rightarrow \infty} \sum_{\nu = -N}^{N} \frac{1}{i\pi (2\nu + 1) - \beta E} = \frac{1}{\exp(\beta E) + 1} - \frac{1}{2},
\label{eq:Ferfunct}
\end{equation}
one obtains from Eq. (\ref {eq:fineqH}) the equation for the unknown function $H$ in its final form,
\begin{equation}
\mid{\bf x}\mid H(\mid{\bf x}\mid) = \frac{e^{2}}{V}\sum_{{\bf k}} \left \{ \frac{1}{\exp \beta({\cal E}_{-}({\bf k}) - \mu) + 1} - \frac{1}{\exp \beta({\cal E}_{+}({\bf k}) - \mu) + 1} \right \} \exp (-i{\bf k}\cdot{\bf x}).
\label{eq:laseqH}
\end{equation}
The equation (\ref {eq:laseqH}) is still a fairly complicated nonlinear integral equation for the
determination of the function $H$; it should be remembered that the $H$-dependence on the right hand
side of Eq. (\ref {eq:laseqH}) occurs through the energy spectra ${\cal E}_{\pm}$ defined in Eq.
(\ref {eq:defcalE}). In the next subsection we consider a limiting case of Eq. (\ref {eq:laseqH}),
for which the solution $H$ can be constructed explicitly.    

\subsection{Solution of the integral equation at T = 0}

In this sub-section we consider a limting case, namely the limit $T \rightarrow 0$, of the integral 
equation (\ref {eq:laseqH}). This limiting case, which corresponds to $\beta \rightarrow \infty$, is  
physically relevant, and furthermore such that one is able to obtain a fairly complete picture of
the solution to Eq. (\ref {eq:laseqH}) in the limit in question.

Let us first note that  Eq. (\ref {eq:laseqH}) is equivalent to the following equation,
\begin{equation}
\int_{V} d^{3}{\bf x}\; \mid{\bf x}\mid H(\mid{\bf x}\mid) \exp (i{\bf k}\cdot{\bf x}) = e^{2} \left \{\frac{1}{\exp \beta({\cal E}_{-}({\bf k}) - \mu) + 1} - \frac{1}{\exp \beta({\cal E}_{+}({\bf k}) - \mu) + 1} \right \}.
\label{eq:FfinH}
\end{equation}
In the limit $\beta \rightarrow \infty$ we have,
\begin{equation}
\lim_{\beta \rightarrow \infty} \frac{1}{\exp\beta(E- \mu) + 1} = \Theta({\cal E}_{F} - E)
\label{eq:stepE}
\end{equation}
where $\Theta$ is the ordinary step function, and we use the notation,
\begin{equation}
\lim_{T \rightarrow 0} \mu =  {\cal E}_{F}
\label{eq:FermE}
\end{equation}

In the limit $T \rightarrow 0$ the analysis of the equation (\ref {eq:FfinH}) boils down to
finding the roots of the following equations,
\begin{equation}
{\cal E}_{\pm}({\bf k}) = {\cal E}_{F}
\label{eq:roots}
\end{equation}
where the quantities ${\cal E}_{\pm}({\bf k})$ have been defined in Eq. (\ref {eq:defcalE}).
 
It turns out that the relevant parameter in the analysis of the Eqns. (\ref {eq:roots}) 
is the value of the quantity $\tilde{H}(0)$, which, without essential loss of generality, can be taken to be 
non-negative, as is easily verified.  One still has to consider separately the cases when ${\cal E}_{F} \geq 0$ 
and ${\cal E}_{F} < 0$, repectively.  The necessity to consider also the case with ${\cal E}_{F} < 0$
is a consequence of the fact that a part of the quasiparticle spectrum ${\cal E}_{-}({\bf k})$ is 
negative. We consider first the case
\begin{equation}
{\cal E}_{F} \geq 0
\label{eq: FermEpos}
\end{equation}

There are then two cases to be considered, namely
\begin{equation}
0 \leq \tilde{H}(0) < {\cal E}_{F} 
\label{eq:H0leqmu}
\end{equation}
and
\begin{equation}
\tilde{H}(0) \geq {\cal E}_{F}
\label{eq:H0geqmu}
\end{equation}

We first consider the case when Eq. (\ref {eq:H0leqmu}) is in force. Then, assuming reasonable 
smoothness of the function $\tilde{H}({\bf k})$, both the equations (\ref {eq:roots}) can have roots
$k = k_{<}$ and $k = k_{>}$, respectively, where
\begin{equation}
{\cal E}_{+}({k_{<}}) \equiv \frac{\hbar^{2}}{2m} \;k_{<}^{2} + \tilde{H}(k_{<}) = {\cal E}_{F}
\label{eq:ksmall}
\end{equation}
and
\begin{equation}
{\cal E}_{-}({k_{>}}) \equiv \frac{\hbar^{2}}{2m} \;k_{>}^{2} - \tilde{H}(k_{>}) = {\cal E}_{F}
\label{eq:klarge}
\end{equation}
In the limit $\beta \rightarrow \infty$  one then gets the following equation from Eq. (\ref {eq:FfinH}), 
\begin{equation}
\int_{0}^{\rightarrow \infty} dr\, r^{2} H(r) \sin(kr) = e^{2} \frac{k}{4\pi}\left \{\Theta(k_{>} - k) - \Theta(k_{<} - k) \right \}
\label{eq:seHtwor}
\end{equation}
From Eq. (\ref {eq:seHtwor}) follows readily that
\begin{equation}
H(r) = \frac{e^{2}}{2\pi^{2} r^{4}}\left \{ [\sin(k_{>}r) - k_{>}r \cos(k_{>}r)] - [\sin(k_{<}r) - k_{<}r \cos(k_{<}r)] \right \},
\label{eq:H(r)etwor}
\end{equation}
from which follows,
\begin{equation}
\tilde{H}(k) = \frac{e^{2}}{\pi} \left \{ k_{>} -  k_{<} + \frac{k^{2} - k_{>}^{2}}{2k}\; \log \left |\frac{k_{>} - k}{k_{>} + k}\right | -  \frac{k^{2} - k_{<}^{2}}{2k}\; \log \left |\frac{k_{<} - k}{k_{<} + k}\right | \right \}.
\label{eq:etildHktwor}
\end{equation}
It then remains to determine the actual values of the roots $k_{<}$ and $k_{>}$ from the equations (\ref {eq:ksmall})
and (\ref {eq:klarge}), respectively. From Eq. (\ref {eq:etildHktwor}) one readily gets the following results,
\begin{equation}
\tilde{H}(k_{<}) = \frac{e^{2}}{\pi} \left \{ k_{>} -  k_{<} - \frac{k_{>}^{2} - k_{<}^{2}}{2k_{<}}\; \log \left |\frac{k_{>} - k_{<}}{k_{>} + k_{<}}\right | \right \} 
\label{eq:etildHksmall}
\end{equation}
and
\begin{equation}
\tilde{H}(k_{>}) = \frac{e^{2}}{\pi} \left \{ k_{>} -  k_{<} - \frac{k_{>}^{2} - k_{<}^{2}}{2k_{>}}\; \log \left |\frac{k_{>} - k_{<}}{k_{>} + k_{<}}\right | \right \} 
\label{eq:etildHklarge}
\end{equation}
Inserting the values for $\tilde{H}(k_{<})$ and $\tilde{H}(k_{>})$ from Eqns. (\ref {eq:etildHksmall}) and
(\ref {eq:etildHklarge}) into the Eqns. (\ref {eq:ksmall}) and  (\ref {eq:klarge}), respectively, one finally 
obtains a pair of transcendental equations for the determination of the roots $k_{<}$ and $k_{>}$. The analysis 
of these equations is facilitated by the introduction of a momentum variable $k_{+}$, related to the Fermi energy
${\cal E}_{F}$ as follows,
\begin{equation}
{\cal E}_{F} \equiv \frac{\hbar^{2}}{2m} k_{+}^{2}
\label {eq:defk+}
\end{equation}
It turns out that the existence of solutions to the transcendental equations (\ref {eq:ksmall}) and  
(\ref {eq:klarge}) determining the roots $k_{<}$ and $k_{>}$, depends on the values of the following 
dimensionless parameter $\alpha_{0}$,
\begin{equation}
\alpha_{0} = \frac{1}{\pi a_{0} k_{+}} 
\label{eq:alfazero}
\end{equation}
where $a_{0}$ denotes the Bohr radius,
\begin{equation}
a_{0} = \frac{\hbar^{2}}{m e^{2}}
\label{eq:Bohrrad}
\end{equation}
It is not difficult to verify that the transcendental equations (\ref {eq:ksmall}) and (\ref {eq:klarge}) have
roots $k_{<}$ and $k_{>}$, with $k_{>} > k_{<} > 0$, if and only if the parameter $\alpha_{0}$ introduced in Eq.
(\ref {eq:alfazero}) satisfies the following inequality,
\begin{equation}
0 < \alpha_{0} < \frac{\sqrt{6}}{12} \approx 0.204.
\label{eq:alfa0twor}
\end{equation}
The inequality (\ref {eq:alfa0twor}) yields a rather large lower bound on the Fermi energy ${\cal E}_{F}$.
Namely, expressing the scale $k_{+}$ in the expression (\ref {eq:defk+}) in terms of the parameter
$\alpha_{0}$ defined in Eq. (\ref {eq:alfazero}), and using the inequality (\ref {eq:alfa0twor})
we obtain,
\begin{equation}
{\cal E}_{F} = \frac{1}{\pi^{2} \alpha_{0}^{2}}\; \frac{e^{2}}{2a_{0}} > \frac{24}{\pi^{2}}\; \frac{e^{2}}{2a_{0}}
\label{eq:lowbFermE}
\end{equation}
The lower bound (\ref {eq:lowbFermE}) is numerically far above the Fermi energies for all existing 
metals. For completeness we nevertheless continue with the analysis of the case at hand. 
 
The pair of transcendental equations (\ref {eq:ksmall}) and (\ref {eq:klarge}) can only be 
solved numerically. For this purpose we introduce dimensionless parameters $u$ and $w$ by writing
\begin{equation}
k_{<} = uk_{+}\, , k_{>} = wk_{+}.
\label{eq:u,w}
\end{equation}
In Table 1 below, we give values of the parameters $u$ and $w$ introduced above in (\ref {eq:u,w}), 
which have been obtained by solving Eqns. (\ref {eq:ksmall}) and (\ref {eq:klarge}) numerically, for a set of 
values of the parameter $\alpha_{0}$ in the interval (\ref {eq:alfa0twor}). 

\begin{table}[h]
\caption{$k_{<} = uk_{+}$, $k_{>} = wk_{+}$}\begin{center}
\begin{tabular}{|c|c|c|c|}
\hline
&$\alpha_{0} = 0.1$ & $\alpha_{0} = 0.15$ & $\alpha_{0} = 0.2$\\
\hline
$u$ & 0.98120 & 0.88503 & 0.61358\\
$w$ & 1.01794 & 1.08974 & 1.19723\\
\hline
\end{tabular}
\end{center}
\end{table}

In view of the condition (\ref {eq:lowbFermE}), the solution discussed above is presumably irrelevant 
for realistic physical systems. Hence we turn our attention to the case when the inequality 
(\ref {eq:H0geqmu}) is in force, which is actually a much simpler case than that already considered.

We now assume that the inequality (\ref {eq:H0geqmu}) is in force. Then, still assuming reasonable 
smoothness of the function $\tilde{H}({\bf k})$, one finds that Eq. (\ref {eq:ksmall}) has no non-negative
root at all. Hence the equation (\ref {eq:FfinH}) can now be written as follows, in the limit
$T \rightarrow \infty$,
\begin{equation}
\int_{0}^{\rightarrow \infty} dr\, r^{2} H(r) \sin(kr) = e^{2} \frac{k}{4\pi} \Theta(k_{>} - k)
\label{eq:sineH}
\end{equation}
where $k_{>}$ is determined by the equation (\ref {eq:klarge}) alone. From Eq. (\ref {eq:sineH}) one
readily obtains the following expression,
\begin{equation}
H(r) = \frac{e^{2}}{2\pi^{2} r^{4}} \left\{\sin(k_{>}r) - k_{>}r \cos(k_{>}r)\right\}.
\label{eq:H(r)}
\end{equation}
which immediately leads to the following expression for the function $\tilde{H}({\bf k})$,
\begin{equation}
\tilde{H}(k) = \frac{e^{2}}{\pi} \left \{ k_{>} + \frac{k^{2} - k_{>}^{2}}{2k}\; \log \left |\frac{k_{>} - k}{k_{>} + k}\right | \right \},
\label{eq:tildeHk}
\end{equation} 
We then finally obtain
\begin{equation}
\tilde{H}(0) = \lim_{k \rightarrow 0} \tilde{H}(k) = \frac{2 e^{2}}{\pi}\;k_{>}
\label{eq:tilde{H}(0)}
\end{equation}
and
\begin{equation}
\tilde{H}(k_{>}) = \frac{ e^{2} }{\pi}\;k_{>}.
\label{eq:H(k_{>})}
\end{equation}
Using the result (\ref {eq:H(k_{>})}) in Eq. (\ref {eq:klarge}) one obtains a simple quadratic equation for
the determination of the root $k_{>}$. It is still expedient to use the momentum scale $k_{+}$ 
defined by Eq. (\ref {eq:defk+}) as well as the dimensionless parameter $\alpha_{0}$ defined by Eq. 
(\ref {eq:alfazero}). Writing
\begin{equation}
k_{>} = wk_{+}
\label{eq:defw}
\end{equation}
one finds that Eq. (\ref {eq:klarge}) is equivalent to the following equation,
\begin{equation}
w^{2} - 2\alpha_{0} w - 1 = 0,
\label{eq:w-eq}
\end{equation}
which gives the final expression for the root $k_{>}$,
\begin{equation}
k_{>} = (\alpha_{0} + \sqrt{\alpha_{0}^{2} + 1})\,k_{+} \equiv (1 + \sqrt{1 + \alpha_{0}^{-2}}) \frac{1}{\pi a_{0}}
\label{eq:rootk2}
\end{equation}

Let us recall that the analysis above is done under the assumption that both the inequalities 
(\ref {eq: FermEpos}) and (\ref {eq:H0geqmu}) are in force. The latter inequality is now equivalent 
to the following condition on the dimensionless parameter $\alpha_{0}$,
\begin{equation}
\alpha_{0} \geq \frac{\sqrt{6}}{12} \approx 0.204
\label{eq:ineqalf}
\end{equation}   
For future reference we note that the inequality (\ref {eq:ineqalf}) implies that
\begin{equation}
\frac{2}{\pi a_{0}} \leq k_{>} \leq \frac{6}{\pi a_{0}}
\label{eq:ineklarge}
\end{equation}

It remains to consider the case of negative Fermi energy ${\cal E}_{F}$,
\begin{equation}
{\cal E}_{F} < 0
\label{eq:FermEneg}
\end{equation}
In this case, there is only one equation for the determination of the root $k_{>}$, namely
Eq. (\ref {eq:klarge}). This equation can have a positive root $k_{>}$ under the condition
(\ref {eq:FermEneg}), if and only if
\begin{equation}
\tilde{H}(0) > - {\cal E}_{F}
\label{eq:negineqH0}
\end{equation}
The analysis proceeds along the same lines as in the previous case with only minor modifications.
We introduce a momentum scale $k_{-}$, determined by the (negative) Fermi energy,  
\begin{equation}
{\cal E}_{F} = - \frac{\hbar^{2}}{2m} k_{-}^{2}
\label{eq:defk-}
\end{equation}
as well as a dimensionless parameter $\gamma_{0}$ analogous to the parameter $\alpha_{0}$ in the
previous case,
\begin{equation}
\gamma_{0} = \frac{1}{\pi a_{0} k_{-}}
\label{eq:defgamma0}
\end{equation}
Writing
\begin{equation}
k_{>} = vk_{-}
\label{eq:defvpar}
\end{equation}
we now obtain the following quadratic equation from Eq. (\ref {eq:klarge}) for the determination of 
the parameter $v$ 
\begin{equation}
v^{2} - 2\gamma_{0}v + 1 = 0
\label{eq:v-eq}
\end{equation}
This gives two roots  $k^{\pm}_{>}$,
\begin{equation}
k^{-}_{>} = (\gamma_{0} - \sqrt{\gamma_{0}^{2} - 1})\,k_{-} \equiv (1 - \sqrt{1 - \gamma_{0}^{-2}}) \frac{1}{\pi a_{0}}
\label{eq:k(-)v2}
\end{equation}
and
\begin{equation}
k^{+}_{>} = (\gamma_{0} + \sqrt{\gamma_{0}^{2} - 1})\,k_{-} \equiv (1 + \sqrt{1 - \gamma_{0}^{-2}}) \frac{1}{\pi a_{0}}
\label{eq:k(+)v2}
\end{equation}
There is the obvious requirement that the roots $k^{\pm}_{>}$ be real; from this follows that we must
necessarily have
\begin{equation}
\gamma_{0}^{2} \geq 1  
\label{eq:cineqga}
\end{equation}
The equations (\ref {eq:H(r)}), (\ref {eq:tildeHk}) and (\ref {eq:tilde{H}(0)}) are now valid as
such, but with $k_{>}$ replaced with either the root given in Eq. (\ref {eq:k(-)v2}) or in 
Eq. (\ref {eq:k(+)v2}). For negative Fermi energy ${\cal E}_{F} < 0$ there are thus two distinct solutions
to the equation (\ref {eq:FfinH}) in the limit $T \rightarrow 0$, for each fixed value of the free 
parameter $\gamma_{0}$ satisfying condition (\ref {eq:cineqga}).

The inequality (\ref {eq:negineqH0}) does not give any additional restrictions on the parameter
$\gamma_{0}$ besides the restriction (\ref {eq:cineqga}). For future reference we finally note that 
the roots $k^{\pm}_{>}$ satisfy the following inequalities, as a consequence of the inequality
(\ref {eq:cineqga}),
\begin{equation}
0 \leq k^{-}_{>} < \frac{1}{\pi a_{0}}
\label{eq: kmin}
\end{equation}
and
\begin{equation}
\frac{1}{\pi a_{0}} \leq k^{+}_{>} < \frac{2}{\pi a_{0}}
\label{eq: kplu}
\end{equation} 
 
Let us summarise the situation so far. We have shown the existence of solutions $H(r)$ to the 
equation (\ref {eq:laseqH}) in the limit $T \rightarrow 0$. If the Fermi energy ${\cal E}_{F}$ is
positive, then the 
solution $H(r)$ takes the form given in Eq. (\ref {eq:H(r)etwor}) if the inequality 
(\ref {eq:H0leqmu})
is in force. This is the case if the dimensionless parameter $\alpha_{0}$ defined by Eq. 
(\ref {eq:alfazero}) satisfies the inequality (\ref{eq:alfa0twor}). If the inequality 
(\ref {eq:H0geqmu})
is in force, then the solution takes the form given in Eq. (\ref{eq:H(r)}). This is the case
if the parameter $\alpha_{0}$ satisfies the inequality (\ref {eq:ineqalf}). The latter solution
can be obtained from the former in the formal limit $k_{<} \rightarrow 0$. In both cases  
the solutions depend on a free parameter $\alpha_{0}$ constrained either by the inequality
(\ref{eq:alfa0twor}) or by the inequality (\ref {eq:ineqalf}). 

If the Fermi energy ${\cal E}_{F}$ is negative, then the solutions $H(r)$ take the form given in
Eq. (\ref{eq:H(r)}),  with either $k^{-}_{>}$ given by Eq. (\ref{eq:k(-)v2})  replacing the
parameter $k_{>}$, or $k^{+}_{>}$ given by Eq. (\ref{eq:k(+)v2}) replacing the parameter $k_{>}$.      

The demonstration given above, is  by no means a replacement of a fullfledged  proof of existence of 
solutions $H(r)$ to the equation (\ref {eq:laseqH}) in the general case, i.e. for nonzero finite $T$.
However, the demonstration above is a proof of existence of solutions $H(r)$ in the limit 
$T \rightarrow 0$, by explicit construction. The solutions obtained will be used below to calculate the
ground state energy of the system under consideration. Before that, we consider in the next section
the effects of fluctuations around the solutions discussed in this subsection.    

\section{Fluctuations}

We have found a stationary point $\overline{B}$ of the effective action (\ref {eq:FinSeff}), 
namely the {\em ansatz} (\ref {eq:Hansatz}), in which the function $H$ is the solution to the 
equation (\ref {eq:laseqH}). We then have to consider the effect of fluctuations around this 
stationary point. We define a fluctuation field ${\cal B}$ as follows,
\begin{equation}
{\cal B}({\bf x}, {\bf x}', \tau)  = B({\bf x}, {\bf x}', \tau) - \overline{B}({\bf x}, {\bf x}', \tau)
\label{eq:calB}    
\end{equation}
Then one has to expand the effective action (\ref {eq:FinSeff}) in powers of the field
${\cal B}$. We will make this expansion up to second order in the field ${\cal B}$. After
some calculations one obtains the dollowing result,
\begin{equation}
S_{eff}[B] = S_{eff}[\overline{B}] + S_{ob}[{\cal B}] + S_{fl}[{\cal B}] - \beta C
\label{eq:finaleff}
\end{equation}
where
\begin{equation}
\label{eq:finCflu}
S_{fl}[{\cal B}]  =  \int_{0}^{\beta} d\tau \int_{0}^{\beta} d\tau ' \int d^{3}{\bf x}\int d^{3}{\bf x}'\int d^{3}{\bf x}''\int d^{3}{\bf x}'''\sum_{\mu, \nu = 0}^{3} \sum_{\alpha \alpha' \alpha'' \alpha'''} R[{\cal B}]
\end{equation}
The quantity $R[{\cal B}]$ in the expression above is given in terms of the Greens functions $G$
defined in Eq. (\ref {eq:defGreen}) as follows, 
\begin{equation}
R[{\cal B}] = - \frac{1}{2}  G_{\alpha\alpha'}({\bf x} - {\bf x}', \tau - \tau'; H) \sigma^{\mu}_{\alpha'\alpha''}{\cal B}_{\mu}({\bf x}', {\bf x}'', \tau') G_{\alpha''\alpha'''}({\bf x}'' - {\bf x}''', \tau' - \tau; H)\sigma^{\nu}_{\alpha'''\alpha}{\cal B}_{\nu}({\bf x}''', {\bf x}, \tau)
\label{eq:Rdef}
\end{equation}

The effective action $S_{eff}[\overline{B}]$ at the stationary point $\overline{B}$ has the 
following explicit form,
\begin{eqnarray}
\label{eq:MN37}
S_{eff}[\overline{B}] & = & \frac{\beta}{2e^{2}} \int d^{3}{\bf x} d^{3}{\bf x}'\mid {\bf x} - {\bf x}'\mid H^{2}({\bf x} - {\bf x}') \\
& - & \sum_{{\bf k}} \log \left \{[1+ e^{-\beta ({\cal E}_{+}({\bf k}) - \mu)}] [1+ e^{-\beta ({\cal E}_{-}({\bf k}) - \mu)}] \right \}\nonumber
\end{eqnarray}

Let us now consider the physical meaning of the stationary point $\overline B$ in some detail.
The effective action $S_{eff}[\overline{B}]$ originates from the following functional integral,
\begin{equation}
\exp\left \{ - S_{eff}[\overline{B}] \right \} = \exp\left \{ - S_{ob}[\overline{B}] \right \} \int {\cal D}[\psi^{*}, \psi] \exp\left \{ - S_{of}[\psi^{*}, \psi] \right \}
\label{eq:MN38}
\end{equation}
where $S_{of}[\psi^{*}, \psi]$ is the following action,
\begin{eqnarray}
\label{eq:MN39}
S_{of}[\psi^{*}, \psi] & = & \int_{0}^{\beta}d\tau \int d^{3} {\bf x} \left \{\psi^{*}({\bf x}, \tau) \frac{\partial}{\partial \tau} \psi({\bf x}, \tau) + \psi^{*}({\bf x}, \tau)\left [ - \frac{\hbar^{2}}{2m} \nabla^{2} - \mu \right ]\psi({\bf x}, \tau) \right \}\\
& + & \int_{0}^{\beta}d\tau \int d^{3} {\bf x}\int d^{3}{\bf x}' H(|{\bf x} - {\bf x}'|)\psi^{*}({\bf x}', \tau)\sigma_{2} \psi({\bf x}, \tau) \nonumber
\end{eqnarray}
The action (\ref {eq:MN39}) describes a system of electrons interacting with a given non-local
magnetic field ${\bf H} = (0, H(|{\bf x}-{\bf x}'|), 0)$. This magnetic field generates 
a correlated creation of an electron with spin up (down) at the point ${\bf x}'$, together with
a simultaneus annihilation of an alectron of opposite spin down (up) at the point ${\bf x}$.

Introducing the following parametrisations for the Grassman variables $\psi$ and $\psi^{*}$,
respectively, in Eq. (\ref {eq:MN38}),
\begin{equation}
\psi({\bf x}, \tau) = \frac{1}{\sqrt{V}}\sum_{\bf k} a_{\bf k}(\tau) e^{i{\bf k}\cdot {\bf x}},\;\psi^{*}({\bf x}, \tau) = \frac{1}{\sqrt{V}}\sum_{\bf k} a^{*}_{\bf k}(\tau) e^{-i{\bf k}\cdot {\bf x}}
\label{eq:MN39.5}
\end{equation}
one can express the action (\ref {eq:MN39}) in terms of the variables $a_{\bf k}(\tau)$ and
$a^{*}_{\bf k}(\tau)$ as follows,
\begin{eqnarray}
\label{eq:MN40}
S_{of}(a^{*}, a) & = & \int_{0}^{\beta} d\tau \sum_{\bf k} \left \{ a^{*}_{\bf k}(\tau)\frac{\partial}{\partial \tau}a_{\bf k}(\tau) + a^{*}_{\bf k}(\tau)[{\cal E}_{\bf k} - \mu] a_{\bf k}(\tau)\right \}\\
& + &  \int_{0}^{\beta} d\tau \sum_{\bf k} \tilde{H}({\bf k}) a^{*}_{\bf k}(\tau)\sigma_{2}a_{\bf k}(\tau) \nonumber
\end{eqnarray}
where $\tilde{H}({\bf k})$ is the Fourier transform of the function $H(|{\bf x}|)$ defined as
in Eq. (\ref {eq:FouriH}).

The Grassmann variables $a^{*}_{{\bf k}Ê\alpha}(\tau)$ and $a_{{\bf k}Ê\alpha}(\tau)$ are 
associated with electron states specified by the wave number ${\bf k}$ and spin $\alpha = \pm$.
In order to diagonalize the action (\ref {eq:MN40}) we introduce new fields $c_{{\bf k}Ê\alpha}$
and $c^{*}_{{\bf k}Ê\alpha}$ by the following unitary transformation,
\begin{eqnarray}
\label{eq:MN41}
& & c_{{\bf k} +} = \frac{1}{\sqrt{2}}(a_{{\bf k} +} - i a_{{\bf k} -})\;, \;c_{{\bf k} -} = \frac{1}{\sqrt{2}}(-ia_{{\bf k} +} + a_{{\bf k} -})Ê\\
& & c^{*}_{{\bf k} +} = \frac{1}{\sqrt{2}}(a^{*}_{{\bf k} +} + i a^{*}_{{\bf k} -})\;, \;c^{*}_{{\bf k} -} = \frac{1}{\sqrt{2}}(ia^{*}_{{\bf k} +} + a^{*}_{{\bf k} -}) \nonumber
\end{eqnarray}
Then the action (\ref {eq:MN40}) can be expressed as follows,
\begin{equation}
\label{eq:MN42}
S_{of}(c^{*},c) = \int_{0}^{\beta}d\tau \sum_{\bf k} \left \{c^{*}_{\bf k}\left [\frac{\partial}{\partial \tau} - \mu \right ] c_{\bf k} + {\cal E}_{+}({\bf k})c^{*}_{{\bf k}+}c_{{\bf k}+} + {\cal E}_{-}({\bf k})c^{*}_{{\bf k}-}c_{{\bf k}-} \right \}
\end{equation}
where the quantities ${\cal E}_{\pm}({\bf k})$ are the energy spectra (\ref {eq:defcalE}) of
the quasiparticles described by the fields (\ref{eq:MN41}). As evident from the relations
(\ref{eq:MN41}), the states of these quasiparticles are special superpositions of electron
states with spin up and spin down. The probability to find an electron in a spin up or down state
is equal to $\frac{1}{2}$ for each quasiparticle spin state. Finally, carrying out the appropriate
functional integration in Eq. (\ref {eq:MN38}) using the form (\ref {eq:MN42}) of the action, one
obtains
\begin{equation}
\exp\left \{-S_{eff}[\overline B] \right \} = \exp\left \{-S_{ob}[\overline B] \right \} \prod_{\bf k}\left \{[1+ e^{-\beta ({\cal E}_{+}({\bf k}) - \mu)}] [1+ e^{-\beta ({\cal E}_{-}({\bf k}) - \mu)}] \right \}
\label{eq:MN43}
\end{equation}
which correponds exactly to the result given in Eq. (\ref {eq:MN37}).

It behooves us still to check whether the self-organised magnetic field
${\bf H} = (0, H(|{\bf x}-{\bf x}'|), 0)$, which is of order $e^{2}$, survives under fluctuations
which are also of order $e^{2}$, or not.  This involves an evaluation of the functional integral
(\ref {eq:Zfin3}), with the effective action given by the second order approximation 
(\ref {eq:finaleff}). One obtains the following result,
\begin{equation}
Z = \exp\left \{-S_{eff}[\overline{B}]\right \} Z_{f\ell}
\label{eq:MNCC45}
\end{equation}
where $S_{eff}[\overline{B}]$ is given by Eq. (\ref {eq:MN37}) and
\begin{equation}
\log Z_{f\ell} = \frac{\beta e^{2}}{4} \int \frac{d^{3}{\bf x}d^{3}{\bf x}'}{|{\bf x}-{\bf x}'|}\, [Tr\,G({\bf x}-{\bf x}', 0+; H)]^{2}
\label{eq:CCMN45a}
\end{equation}
In Eq. (\ref {eq:CCMN45a}) above, the notation $Tr$ stands for the appropriate spin index sums 
of the indicated quantities involving the Green's functions defined by Eq. (\ref {eq:defGreen}).
The term $\beta C$ in Eq. (\ref {eq:finaleff}) was exactly cancelled by the direct Coulomb term
which appeared in the average value $<S_{f\ell}({\cal B})>_{0}$.

Using the results (\ref {eq:MNCC45}) and (\ref {eq:CCMN45a}) on gets the grand canonical
potential $\Omega$,
\begin{equation}
\Omega = -k_{B} T \log Z
\label{eq:CCOmega}
\end{equation}
from which one obtains the energy $E_{mag}$ of the electron system with the self-organized
magnetic field ${\bf H}$,
\begin{eqnarray}
\label{eq:MN46}
E_{mag} & = & \frac{1}{2e^{2}} \int d^{3}{\bf x} d^{3}{\bf x}'\mid {\bf x} - {\bf x}'\mid H^{2}({\bf x} - {\bf x}')\\
& + &\sum_{{\bf k}}\left \{{\cal E}_{+}({\bf k})f({\cal E}_{+}({\bf k})) + {\cal E}_{-}({\bf k})f({\cal E}_{-}({\bf k}))\right \} \nonumber \\
& - & \frac{e^{2}}{4}\int \frac{d^{3}{\bf x}d^{3}{\bf x}'}{|{\bf x}-{\bf x}'|} \,[Tr\,G({\bf x}-{\bf x}', 0+; H)]^{2} \nonumber
\end{eqnarray}
where the function $f({\cal E})$ denotes the Fermi function,
\begin{equation}
\label{eq:Fermif}
f({\cal E}) = \frac{1}{e^{\beta ({\cal E} - \mu)} + 1}
\end{equation}

In the expression (\ref {eq:MN46}) the first term corresponds to the energy of the 
self-organised magnetic field ${\bf H}$,and the second term corresponds to the energy of 
electrons interacting with this self-organised magnetic field. The last term corresponds 
to the exchange energy and to fluctuations about the self-organised magnetic field (mean field)
calculated to the order of $e^{2}$. This term does not dominate over the two first terms in
Eq. (\ref {eq:MN46}). Hence the fluctuations do not destroy the effects of the self-organised
magnetic field. 

\section{Ground state energy}

In the case $T = 0$ the function $H(r)$, which determines the self-organised magnetic field
is known explicitly, as shown in subsection (3.4.). If the inequality (\ref {eq:H0leqmu}) is in force,
then the function $H(r)$ is given by Eq. (\ref{eq:H(r)etwor}). If the follwing inequality holds,
\begin{equation}
\tilde{H}(0) \geq |{\cal E}_{F}|
\label{eq:geninEF}
\end{equation}
then the function $H(r)$ is given by Eq. (\ref{eq:H(r)}). It should be noted that the latter case
is obtained from the former in the formal limit $k_{<} \rightarrow 0$, whence we continue our
analysis using the expression (\ref{eq:H(r)etwor}). Using this expression, or rather
its Fourier transform (\ref {eq:etildHktwor}), we can evaluate the energy $E_{mag}$ in Eq. 
(\ref {eq:MN46}) in a fairly explicit form. After straightforward calculations one obtains,
\begin{eqnarray}
\label{eq:MN47}
E_{mag} & = & \frac{V}{2\pi^{2}} \left \{\frac{\hbar^{2}}{10m}(k_{>}^{5} + k_{<}^{5})\right \}\\
& - & \frac{V}{2\pi^{2}} \left \{\frac{e^{2}}{8\pi}\left [3(k_{>}^{4} + k_{<}^{4}) - k_{>}k_{<}(k_{>}^{2} + k_{<}^{2}) + 6(k_{>}^{2} - k_{<}^{2})^{2} \log \frac{k_{>} + k_{<}}{k_{>} - k_{<}}\right ] \right \}\nonumber
\end{eqnarray}

Before proceeding further, let us recall that the equations (\ref {eq:klarge}) and (\ref {eq:ksmall})
have two positive roots $k_{>}$ and $k_{<}$ if and only if the inequality (\ref {eq:alfa0twor})
is in force. This inequality in turn implies the lower bound (\ref{eq:lowbFermE}}) for the Fermi energy
${\cal E}_{F}$, which numerically is far above the Fermi energies for all known metals. 

The physically relevant case obtains when the inequality (\ref {eq:geninEF}) holds. The corresponding
formula for $E_{mag}$ can be obtained by passing to the formal limit $k_{<} \rightarrow 0$ in Eq. 
(\ref {eq:MN47}). Thus, if condition (\ref {eq:geninEF}) holds true, then
\begin{equation}
E_{0mag} \equiv \lim_{k_{<} \rightarrow 0} E_{mag} = \frac{V}{2\pi^{2}} \left \{\frac{\hbar^{2}}{10m}k_{>}^{5} - \frac{3e^{2}}{8\pi}k_{>}^{4} \right \}
\label{eq:E0mag}
\end{equation}
where the quantity $k_{>}$ is given either by the equation (\ref{eq:rootk2}), which implies that 
$k_{>}$ is in the interval (\ref {eq:ineklarge}) or by one or the other of the equations 
(\ref {eq:k(-)v2}) or (\ref {eq:k(+)v2}), which restrict the quantity $k_{>}$ to be in one or the other
of the intervals (\ref {eq: kmin}) or (\ref {eq: kplu}).

Thus, for realistic metals, we expect the ground state energy to be given by Eq. (\ref {eq:E0mag}),
which originates from quasiparticles in the spin down states only, with energy spectrum 
${\cal E}_{-}({\bf k})$.

At this point it is appropriate to recall that we are supposed to calculate the energy per particle.
The number of electrons is given as follows,
\begin{eqnarray}
\label{eq:numberN}
N & \equiv & \sum_{\alpha} \int d^{3}{\bf x}<\hat{\psi}_{\alpha}^{\dagger}({\bf x})\hat{\psi}_{\alpha}({\bf x})>\\
& = & \frac{V}{(2\pi)^{3}} \int d^{3} {\bf k} \left [f({\cal E}_{-}({\bf k})) + f({\cal E}_{+}({\bf k}))\right ]\nonumber
\end{eqnarray}
where the functions $f({\cal E}_{\pm})$ denote the appropriate Fermi functions (\ref {eq:Fermif}).
 
In the limit $T \rightarrow 0$ the integral (\ref{eq:numberN}) can immediately be evaluated by use
of the formula (\ref {eq:stepE}), with the result,
\begin{equation}
N = \frac{V}{6\pi^{2}} (k_{>}^{3} + k_{<}^{3})
\label{eq:density}
\end{equation}
Thus the parameters $k_{>}$ and $k_{<}$ are related to the average particle density as indicated 
above. In particular, in the case $k_{<} \rightarrow 0$ we obtain the following relation 
from Eq. (\ref {eq:density}),
\begin{equation}
\frac{V}{N} = 6\pi^{2}\,k_{>}^{-3}
\label{eq:VNk>rel}
\end{equation}
Using the reult (\ref {eq:VNk>rel}) in Eq. (\ref {eq:E0mag}) we obtain the energy per particle,
\begin{equation}
\frac{E_{0mag}}{N} = \frac{3}{2} e^{2} \left \{ \frac{a_{0}}{5} k_{>}^{2} - \frac{3}{4\pi} k_{>} \right \}
\label{eq:finE0k}
\end{equation}
To obtain the genuine ground state energy, we still have to minimise the expression (\ref {eq:finE0k})
with respect to the parameter $k_{>}$, which can vary within certain limits, as discussed above.
One readily finds that the expression (\ref {eq:finE0k}) has a minumum for the following value of
$k_{>}$,
\begin{equation}
(k_{>})_{min} = \frac{15}{8} \, \frac{1}{\pi a_{0}} = \frac{1.875}{\pi a_{0}}
\label{eq:Ckmin}
\end{equation}
The corresponding minimum value of the energy per particle is,
\begin{equation}
\left (\frac{E_{0mag}}{N} \right )_{min} = - 15 \left (\frac{3}{8\pi} \right )^{2} \, \frac{e^{2}}{2a_{0}} \approx -0.214\, \frac{e^{2}}{2a_{0}}
\label{eq:E0magmin}
\end{equation}
Numerically the result (\ref {eq:E0magmin}) corresponds to the following value,
\begin{equation}
\left (\frac{E_{0mag}}{N} \right )_{min} \approx - 2.91\; eV
\label{eq:numEmin}
\end{equation}

The minimum (\ref {eq:E0magmin}) occurs at the value (\ref {eq:Ckmin}) for the parameter $k_{>}$,
which is within the limits in Eq. (\ref {eq: kplu}). Hence the minimum considered here corresponds
to a {\em negative} value for the Fermi energy ${\cal E}_{F}$,
\begin{equation}
{\cal E}_{F} < 0
\label{eq:negEF}
\end{equation}

In conclusion we wish to compare the results given above, with the ground state energy $E_{0}$ of the 
unmagnetised case considered in Sec. 2.  

The ground state energy in question is obtained by adding the quantities $E_{kin}$ and $E_{exch}$ 
given by Eq.(\ref {eq:Ekin}) and Eq. (\ref {eq:Eexch}), respectively,
\begin{equation}
\label{eq:MN49}
E_{0} = \frac{e^{2} N}{2a_{0}}\left \{\frac{3}{5}\left (\frac{9\pi}{4}\right )^{\frac{2}{3}} \frac{1}{r_{s}^{2}} - \frac{3}{2\pi}\left (\frac{9\pi}{4}\right )^{\frac{1}{3}} \frac{1}{r_{s}} \right \} \approx \frac{e^{2}N}{2a_{0}}\left \{\frac{2.21}{r_{s}^{2}} - \frac{0.916}{r_{s}} \right \}
\end{equation}
The parameter $r_{s}$ occuring in the equation (\ref {eq:MN49}) above, is related to the volume per
electron, as detailed in Sec. 2, i.e.
\begin{equation}
r_{s} = \frac{r_{0}}{a_{0}}
\label{eq:defress}
\end{equation}
and
\begin{equation}
\frac{4}{3} \pi r_{0}^{3} = \frac{V}{N}
\label{eq:defr0}
\end{equation}
Thus, using Eq. (\ref {eq:VNk>rel}) and the equations (\ref {eq:defress}) and (\ref {eq:defr0})
above, we obtain a relation between the parameters $k_{>}$ and $r_{s}$,
\begin{equation}
r_{s} k_{>} a_{0} = \left (\frac{9\pi}{2} \right )^{\frac{1}{3}}
\label{eq:resska}
\end{equation}
Eliminating the parameter $k_{>}$ from Eq. (\ref {eq:finE0k}) by means of Eq. (\ref {eq:resska})
above, one obtains,
\begin{equation}
E_{0mag} = \frac{e^{2}N}{2a_{0}} \left \{ \frac{3}{5}\left(\frac{9\pi}{2}\right )^{\frac{2}{3}} \frac{1}{r_{s}^{2}} - \frac{3}{2\pi}\left(\frac{9\pi}{2}\right )^{\frac{1}{3}} \frac{3}{2r_{s}} \right \} \approx \frac{e^{2}N}{2a_{0}} \left \{ \frac{3.508}{r_{s}^{2}} - \frac{1.731}{r_{s}} \right \}
\label{eq:MN50}
\end{equation}

The energy per particle of the magnetically ordered state  is smaller than the corresponding
energy per particle in the unmagnetised state, provided the parameter $r_{s}$ exceeds a certain limit
$r_{sc}$,
\begin{equation}
\frac{E_{0mag}}{N}  <  \frac{E_{0}}{N} \Longleftrightarrow r_{s} > r_{sc}
\label{eq:critrs}
\end{equation}
A comparison of the two expressions (\ref {eq:MN49}) and (\ref {eq:MN50}) above, yields a numerical
estimate for the lower limit $r_{sc}$ in (\ref {eq:critrs}),
\begin{equation}
r_{sc} \approx 1.592
\label{eq:numrcrits}
\end{equation}

The condition  $r_{s} > 1.592$ is satisfied for all metals, so that the
energy per particle in the magnetically ordered state is smaller than the energy per particle in the 
unmagnetised state in all the relevant cases.

Let us finally minimise the energy per particle (\ref {eq:MN49}) for the unmagnetised state with respect
to $r_{s}$. One obtains,
\begin{equation}
(r_{s})_{min} \approx 4.83\;, \; \left (\frac{E_{0}}{N}\right )_{min} \approx - 1.29\; eV
\label{eq:MN52.5}
\end{equation}
The minimum value $- 2.91\; eV$ for the energy per particle in the magnetically ordered state,
which was obtained above in Eq (\ref {eq:E0magmin}), is much closer to the experimental values for 
most metals \cite{Metals} than the value $-1.29\; eV$ obtained above from the energy per particle in 
the unmagnetised case. The minimum in question occurs at
\begin{equation}
(r_{s})_{min} = \frac{8\pi}{15} \left (\frac{9\pi}{2} \right )^{\frac{1}{3}} \approx 4.05
\label{eq:last}
\end{equation}
as is readily found from Eqns. (\ref {eq:Ckmin}) and (\ref {eq:resska}).

In fact, the binding energies of all metals exceed numerically considerably the value $1.29\; eV$, 
and are closer to the value $2.91\; eV$ obtained for the magnetised case considered here.

\section{Summary and conclusions}

Starting form the simplest theory of metals, we have derived a particular mechanism by means of
which the Coulomb interaction between electrons can generate a new kind of magnetically ordered 
state in a metal. The self-organised magnetic field geberated by this mechanism is represented by
an order parameter which is a non-local field of a special kind. The self-organised field does 
not produce spin polarisations of the electrons, but generates particular long range correlations
between electrons of opposite spin in different points ${\bf x}$ and ${\bf x}'$. The electron
plasma in the model is described in terms of interacting quasiparticles of fermionic type. Each
quasiparticle state is shown to be a special superposition of an electron spin up and spin down state,
such that the probability of finding an electron with either spin up or spin down in each 
quasiparticle state equals $\frac{1}{2}$. Hence the spontaneously organised magnetic field
does not generate polarisations of the electrons in the metal but only polarisations of the
quasiparticles.

The ground state of the electron plasma has been shown to be described by quasiparticles with
a specific absolute spin-polarisation, but with no spin-polarisations of the electrons, for
such values of e.g. the Fermi energy, which are valid for all existing metals. Hence the 
ground state in the simplest model of metals is not a ferromagnetic state as advocated by
Bloch \cite{Bloch}, but rather a state involving the spontaneously generated non-local magnetic
field derived in this paper. The ground state properties of this magnetised state give rise to
theoretical numerical values for the average distance between electrons and for the binding 
energy which are in reasonable agreement with experimetal data.

It ought to be possible to use our theoretical analysis as a starting point for experimental
research in condensed matter physics aiming at observing and studying the magnetically ordered
states predicted in this paper.

\vfill
 
{\bf Acknowledgements}

A large portion of the work described in this paper was done while one of the authors (M. N.)
was a summer visitor at the Helsinki Institute of Physics (HIP) in Helsinki. The hospitality of
HIP, in particular T. Ala-Nissil{\"a}, is gratefully acknowledged.

\vfill\eject

\end{document}